\documentclass[12pt,a4paper]{article}

\usepackage{amssymb,amsmath,eucal}
\usepackage[dvips]{lscape,graphicx}
\usepackage{cite}

\renewcommand{\thefootnote}{\arabic{footnote}}

\newcommand{\bc}{\begin{center}}
\newcommand{\ec}{\end{center}}
\newcommand{\bd}{\begin{displaymath}}
\newcommand{\ed}{\end{displaymath}}
\newcommand{\be}{\begin{equation}}
\newcommand{\ee}{\end{equation}}
\newcommand{\ba}{\begin{array}}
\newcommand{\ea}{\end{array}}
\newcommand{\bt}{\begin{tabular}}
\newcommand{\et}{\end{tabular}}

\newcommand{\ov}{\overline}
\newcommand{\ds}{\displaystyle}

\newcommand{\ct}{\cite}

\newcommand{\lb}{\label}
\newcommand{\bp}{\begin{picture}}
\newcommand{\ep}{\end{picture}}
\newcommand{\bfi}{\begin{figure}}
\begin{document}

\title{\huge\bf {Phase Transition in
Gauge Theories and Multiple Point Model}}

\author{\\{\bf L.V.Laperashvili\footnotemark[1]}\\[0.3cm]
{\bf H.B.Nielsen{\renewcommand{\thefootnote}{\arabic{footnote})}\footnotemark[1]}\;\,
\footnotemark[7]}\\[0.2cm]
{\bf D.A.Ryzhikh{\renewcommand{\thefootnote}{\arabic{footnote})}\footnotemark[2]}\;\;\;%
\footnotemark[1]\:\,\footnotemark[1]\:\,\footnotemark[1]}\\[1cm]
\it Institute of Theoretical and Experimental Physics,\\
\it Moscow, Russia}

\date{}

{\renewcommand{\thefootnote}{\arabic{footnote})}
\addtocounter{footnote}{+1}
\footnotetext{Niels Bohr Institute, Copenhagen, Denmark.}}

{\renewcommand{\thefootnote}{\arabic{footnote})}
\addtocounter{footnote}{+1}
\footnotetext{ Institute of Theoretical and Experimental Physics, Moscow, Russia.}}

{\renewcommand{\thefootnote}{\fnsymbol{footnote}}
\addtocounter{footnote}{-1}
\footnotetext{{\bf E-mail}: laper@heron.itep.ru}}

{\renewcommand{\thefootnote}{\fnsymbol{footnote}}
\addtocounter{footnote}{+6}
\footnotetext{{\bf E-mail}: hbech@alf.nbi.dk}}

{\renewcommand{\thefootnote}{\fnsymbol{footnote}$*$}
\footnotetext{{\bf E-mail}: ryzhikh@heron.itep.ru}}

\maketitle

\thispagestyle{empty}

\newpage


In the present paper the phase transition in the regularized $U(1$) gauge theory
is investigated using the dual Abelian Higgs model of scalar monopoles. The
corresponding renormalization group improved effective potential, analogous to
the Coleman--Weinberg's one, was considered in the two-loop approximation for $\beta$
functions, and the phase transition (critical) dual and non-dual
couplings were calculated in the $U(1)$ gauge theory. It was shown that the
critical value of the renormalized electric fine structure constant
$\alpha_{\text{crit}}\approx 0.208$ obtained in this paper coincides with the
lattice result for compact QED: $\alpha_{\text{crit}}^{\text{lat}}\approx 0.20\pm 0.015$.
This result and the behavior of $\alpha$ in the vicinity of the phase
transition point were compared with the Multiple Point Model prediction
for the values of $\alpha$ near the Planck scale. Such a comparison is very
encouraging for the Multiple Point Model assuming the existence of the
multiple critical point at the Planck scale.


\thispagestyle{empty}

\newpage

\pagenumbering{arabic}

\section{Introduction}

The philosophy of the Multiple Point Model (MPM) suggested in \ct{1y}
and developed in [2--4] leads to the necessity to investigate
the phase transition in different gauge theories. According to MPM,
there is a special point -- the Multiple Critical Point (MCP) -- on the
phase diagram of the fundamental regularized gauge theory $G$, which is
a point where the vacua of all fields existing in Nature are degenerate,
having the same vacuum energy density. Such a phase diagram has axes given
by all coupling constants considered in theory.
MPM assumes the existence of MCP at the Planck scale.

A lattice model of gauge theories is the most convenient formalism for the
realization of the MPM ideas. In the simplest case we can imagine our
space-time as a regular hypercubic (3+1)-lattice with the parameter $a$
equal to the fundamental (Planck) scale: $a = \lambda_{\text{P}} = 1/M_{\text{Pl}}$, where
\begin{equation}
M_{\text{Pl}}=1.22\times 10^{19}\, {\mbox{GeV}}.
\lb{1a}
\end{equation}
Lattice gauge theories, first introduced by Wilson \ct{1s} for studying
the problem of confinement, are described by the following simplest action:
\begin{equation}
S = - \frac{\beta}{N}\sum_{\text{p}} \text{Re}(\text{Tr}\,{\cal U}_{\text{p}}),
\lb{1}
\end{equation}
where the sum runs over all plaquettes of a hypercubic lattice and
${\cal U}_{\text{p}}$ is the product around the plaquette $p$ of the link
variables in the $N$-dimensional fundamental representation of the
gauge group $G$; $\beta=1/g_0^2$ is the lattice constant and $g_0$ is the bare coupling constant of the gauge
theory considered. Monte Carlo simulations of these simple Wilson lattice
theories in the four dimensions showed a (or an almost) second-order
deconfining phase transition for $U(1)$ \ct{2s,3s}, a crossover
behavior for $SU(2)$ and $SU(3)$ \ct{4s,5s}, and a first-order
phase transition for $SU(N)$ with $N\ge 4$ \ct{6s}. Bhanot and
Creutz \ct{7s,8s} have generalized the simple Wilson action,
introducing two parameters in action:
\begin{equation}
S = \sum_p[-\frac{\beta_{\text{f}}}{N}\text{Re}(\text{Tr}\,{\cal U}_{\text{p}}) -
\frac{\beta_{\text{A}}}{N^2-1}\text{Re}(\text{Tr}_{\text{A}}{\cal U}_{\text{p}})],   \lb{2}
\end{equation}
where $\beta_{\text{f}}$, $\text{Tr}$ and $\beta_{\text{A}}$, ${\text{Tr}}_{\text{A}}$ are  respectively the lattice constants and traces in
the fundamental and adjoint representations of $SU(N)$ considered in this action for ${\cal U}_{\text{p}}$.
The phase diagrams, obtained for the generalized lattice $SU(2)$ and
$SU(3)$ theories (\ref{2}) by Monte Carlo methods in \ct{7s,8s},
showed the existence of a triple point which is a boundary point of three
first-order phase transitions: the "Coulomb-like" and confining $SU(N)/Z_{\text{N}}$, $Z_{\text{N}}$ phases meet together
at this point. From the triple point emanate three phase border lines which separate the corresponding
phases. The $Z_{\text{N}}$ phase transition is a "discreteness" transition, occurring
when lattice plaquettes jump from the identity to nearby elements
in the group. The $SU(N)/Z_{\text{N}}$ phase transition is due to a condensation
of monopoles (a consequence of the non-trivial $\Pi_1$ of the group).

Monte Carlo simulations of the $U(1)$ gauge theory, described by the two-parameter lattice action \ct{9s,10s}:
\begin{equation}
S = \sum_{\text{p}}[\beta^{\text{lat}} \cos \Theta_{\text{p}} + \gamma^{\text{lat}} \cos2\Theta_{\text{p}}],\quad
{\mbox {where}} \quad {\cal U}_{\text{p}} = e^{i\Theta_{\text{p}}},     \lb{3}
\end{equation}
also indicate the existence of a triple point on the corresponding
phase diagram (see Fig.1): "Coulomb-like", totally confining, and $Z_2$
confining phases come together at the triple point shown in Fig.1.

Monte Carlo simulations of the lattice $U(1)$
gauge theory, described by the simple Wilson action corresponding
to the case $\gamma^{\text{lat}}=0$ in Eq.(\ref{3}), give us \ct{10s}:
\begin{equation}
\alpha_{\text{crit}}^{\text{lat}}\approx 0.20\pm 0.015\quad
{\mbox{and}} \quad {\tilde \alpha}_{\text{crit}}^{\text{lat}}\approx 1.25\pm 0.10,
                                                            \lb{4}
\end{equation}
where $\alpha = e^2/4\pi $ and ${\tilde \alpha}=g^2/4\pi$ are the electric
and magnetic fine structure constants, containing the electric charge
$e$ and magnetic charge $g$, respectively.
The lattice artifact monopoles are responsible for the confinement
mechanism in lattice gauge theories what is confirmed by many
numerical and theoretical investigations (see reviews \ct{11s} and papers
\ct{12s}). The simplest effective dynamics describing the
confinement mechanism in the pure gauge lattice $U(1)$ theory
is the dual Abelian Higgs model of scalar monopoles \ct{13s}.

In our previous papers [1--3] the calculations of the $U(1)$
phase transition (critical) coupling constant were connected with the
existence of artifact monopoles in the lattice gauge theory and also
in the Wilson loop action model \ct{3y}.
Here we consider the Higgs Monopole Model (HMM) approximating
the lattice artifact monopoles as fundamental pointlike
particles described by the Higgs scalar field.
The phase border separating the Coulomb-like and confinement phases
is investigated by the method developed in MPM, where degenerate vacua
are considered. The phase transition Coulomb--confinement is given
by the condition when the first local minimum of the effective potential
 is degenerate with its
second minimum.
Considering the renormalization group improvement of the effective
Coleman--Weinberg potential \ct{20s,21s} written for the dual sector of
scalar electrodynamics in the two-loop approximation, we have calculated the
$U(1)$ critical values of the magnetic fine structure constant ${\tilde
\alpha}_{\text{crit}} = g^2_{\text{crit}}/4\pi\approx 1.20$ and electric fine structure
constant $\alpha_{\text{crit}} = \pi/g^2_{\text{crit}}\approx 0.208 $ (by the Dirac
relation). These values coincide with the lattice result (\ref{4}).

Investigating the phase transition in HMM
we have pursued two objects. From one side, we had an aim to
explain the lattice results. But we had also another aim.

According to MPM, at the Planck scale there exists a multiple critical point,
which is a boundary point of the phase transitions in $U(1)$, $SU(2)$, and $SU(3)$
sectors of the fundamental regularized gauge theory $G$.
The idea of \ct{1y} was that the corresponding critical couplings
coincide with the lattice ones. Our calculations in HMM indicate that the
Higgs scalar monopole fields are responsible for the phase transition
Coulomb--confinement, giving the same lattice values of critical
couplings.
By this reason, the results
of the present paper are very encouraging for the Anti-Grand Unification
Theory (AGUT) [\citen{33}--\citen{38}], which was developed previously as a realistic
alternative to SUSY GUTs.  This paper is also devoted to the discussion of the
problems of AGUT, which is used in conjunction with MPM.

\section{The Coleman--Weinberg Effective Potential for the Higgs Monopole Model}

As it was mentioned in Introduction, the dual Abelian Higgs model of scalar
monopoles (shortly HMM) describes the dynamics of confinement in lattice
theories. This model, first suggested in \ct{13s}, considers the
following Lagrangian:
\begin{equation}
    L = - \frac{1}{4g^2} F_{\mu\nu}^2(B) + \frac{1}{2} |(\partial_{\mu} -
           iB_{\mu})\Phi|^2 - U(\Phi),\quad              \lb{5}
{\mbox{where}}\quad
 U(\Phi) = \frac{1}{2}\mu^2 {|\Phi|}^2 + \frac{\lambda}{4}{|\Phi|}^4
\end{equation}
is the Higgs potential of scalar monopoles with magnetic charge $g$, and
$B_{\mu}$ is the dual gauge (photon) field interacting with the scalar
monopole field $\Phi$.  In this model $\lambda$ is the self-interaction constant of scalar fields, and the mass parameter
$\mu^2$ is negative. In Eq.(\ref{5}) the complex scalar field $\Phi$ contains
the Higgs ($\phi$) and Goldstone ($\chi$) boson fields:
\begin{equation}
          \Phi = \phi + i\chi.             \lb{7}
\end{equation}

The effective potential in the Higgs scalar electrodynamics (HSED)
was first calculated by Coleman and Weinberg \ct{20s} in the one-loop
approximation. The general method of its calculation is given in the
review \ct{21s}. Using this method, we can construct the effective potential
for HMM. In this case the total field system of the gauge ($B_{\mu}$)
and magnetically charged ($\Phi$) fields is described by
the partition function which has the following form in Euclidean space:
\begin{equation}
      Z = \int [DB][D\Phi][D\Phi^{+}]\,e^{-S},     \lb{8}
\end{equation}
where the action $S = \int d^4x L(x) + S_{\text{gf}}$ contains the Lagrangian
(\ref{5}) written in Euclidean space and gauge fixing action $S_{\text{gf}}$.
Let us consider now a shift
\begin{equation}
\Phi (x) = \Phi_{\text{b}} + {\hat\Phi}(x)
\lb{9}
\end{equation}
with $\Phi_{\text{b}}$ as a background field and calculate the
following expression for the partition function in the one-loop
approximation:
$$
  Z = \int [DB][D\hat \Phi][D{\hat \Phi}^{+}]
   \exp\{ - S(B,\Phi_{\text{b}})
   - \int d^4x [\frac{\delta S(\Phi)}{\delta \Phi(x)}|_{\Phi=
   \Phi_{\text{b}}}{\hat \Phi}(x) + \text{h.c.} ]\}=\\
$$
\begin{equation}
    =\exp\{ - F(\Phi_{\text{b}}, g^2, \mu^2, \lambda)\}.      \lb{10}      
\end{equation}
Using the representation (\ref{7}), we obtain the effective potential:
\begin{equation}
  V_{\text{eff}} = F(\phi_{\text{b}}, g^2, \mu^2, \lambda)        \lb{11}
\end{equation}
given by the function $F$ of Eq.(\ref{10}) for the constant background
field $ \Phi_{\text{b}} = \phi_{\text{b}} = \mbox{const}$. In this case the one-loop
effective potential for monopoles coincides with the expression
of the effective potential calculated by the authors of \ct{20s} for HSED and extended to the massive theory
(see review \ct{21s}):
$$
 V_{\text{eff}}(\phi_{\text{b}}^2) = \frac{\mu^2}{2} {\phi_{\text{b}}}^2 +
                 \frac{\lambda}{4} {\phi_{\text{b}}}^4
  + \frac{1}{64\pi^2}[ 3g^4 {\phi_{\text{b}}}^4\log\frac{\phi_{\text{b}}^2}{M^2} +
$$
\begin{equation}
   {(\mu^2 + 3\lambda {\phi_{\text{b}}}^2)}^2\log\frac{\mu^2 + 3\lambda\phi_{\text b}^2}{M^2}
 + {(\mu^2 +\lambda \phi_{\text{b}}^2)}^2\log\frac{\mu^2
 + \lambda \phi_{\text{b}}^2}{M^2}] + C,
                           \lb{12}
\end{equation}
where $M$ is the cut-off scale and $C$ is a constant not depending on
$\phi_{\text{b}}^2$.

The effective potential (\ref{11}) has several minima. Their position
depends on $g^2, \mu^2,$ and $\lambda$.
If the first local minimum occurs at $\phi_{\text{b}}=0$ and $V_{\text{eff}}(0)=0$,
it corresponds to the so-called symmetrical phase, which is
the Coulomb-like phase in our description. Then it is easy to determine
the constant $C$ in Eq.(\ref{12}):
\begin{equation}
        C = - \frac{\mu^4}{16{\pi^2}}\log \frac {\mu}M,  \lb{13}
\end{equation}
and we have the effective potential for HMM described
by the following expression:
\begin{equation}
V_{\text{eff}}(\phi^2_{\text{b}})
= \frac{\mu^2_{\text{run}}}{2}\phi_{\text{b}}^2 + \frac{\lambda_{\text{run}}}{4}\phi_{\text{b}}^4
     + \frac{\mu^4}{64\pi^2}\log\frac{(\mu^2 + 3\lambda \phi_{\text{b}}^2)(\mu^2 +
       \lambda \phi_{\text{b}}^2)}{\mu^4}.
                                                \lb{14}
\end{equation}
Here $\lambda_{\text{run}}$ is the running self-interaction constant
given by Eq.(\ref{12}):
\begin{equation}
  \lambda_{\text{run}}(\phi_{\text{b}}^2)
   = \lambda + \frac{1}{16\pi^2} [ 3g^4\log \frac{\phi_{\text{b}}^2}{M^2}
   + 9{\lambda}^2\log\frac{\mu^2 + 3\lambda \phi_{\text{b}}^2}{M^2} +
     {\lambda}^2\log\frac{\mu^2 + \lambda\phi_{\text{b}}^2}{M^2}].   \lb{15}
\end{equation}
The running squared mass of the Higgs scalar monopoles also follows from
Eq.(\ref{12}):
\begin{equation}
   \mu^2_{\text{run}}(\phi_{\text{b}}^2)
   = \mu^2 + \frac{\lambda\mu^2}{16\pi^2}[ 3\log\frac{\mu^2 +
   3\lambda \phi_{\text{b}}^2}{M^2} + \log\frac{\mu^2 + \lambda\phi_{\text{b}}^2}{M^2}].
                                  \lb{16}
\end{equation}

\section{Renormalization group equations in the Higgs monopole model}

The renormalization group equations (RGE) for the effective potential means that the potential cannot
depend on a change in the arbitrary parameter -- renormalization scale $M$,
i.e., $dV_{\text{eff}}/dM = 0$. The effects of changing it are absorbed into
changes in the coupling constants, masses, and fields, giving so-called
running quantities.

Considering the renormalization group (RG) improvement of the effective potential \ct{20s,21s}
and choosing the evolution variable as
\begin{equation}
                   t = \log(\phi^2/M^2),     \lb{18} 
\end{equation}
we have the following RGE for the improved $V_{\text{eff}}(\phi^2)$
with $\phi^2\equiv \phi^2_{\text{b}}$ \ct{22s}:
\begin{equation}
   (M^2\frac{\partial}{\partial M^2} + \beta_{\lambda}\frac{\partial}
{\partial\lambda} + \beta_{g}\frac{\partial}{\partial g^2} +
\beta_{(\mu^2)}\mu^2\frac{\partial}{\partial \mu^2} - \gamma\phi^2
\frac{\partial}{\partial \phi^2})V_{\text{eff}}(\phi^2) = 0,    \lb{19} 
\end{equation}
where $\gamma$ is the anomalous dimension and $\beta_{(\mu^2)}$,
$\beta_{\lambda}$, and $\beta_{g}$ are the RG $\beta$ functions for mass,
scalar, and gauge couplings, respectively. RGE (\ref{19}) leads to the
following form of the improved effective potential \ct{20s}:
\begin{equation}
     V_{\text{eff}} = \frac{1}{2}\mu^2_{\text{run}}(t)G^2(t)\phi^2 +
                 \frac{1}{4}\lambda_{\text{run}}(t)G^4(t)\phi^4.  \lb{20} 
\end{equation}
In our case:
\begin{equation}
 G(t) = \exp[-\frac{1}{2}\int_0^t dt'\,\gamma\left(g_{\text{run}}(t'),
         \lambda_{\text{run}}(t')\right)].                         \lb{21}  
\end{equation}
A set of ordinary differential equations (RGE) corresponds to Eq.(\ref{19}):

\begin{equation}
    \frac{d\lambda_{\text{run}}}{dt} = \beta_{\lambda}\left(g_{\text{run}}(t),\,
                    \lambda_{\text{run}}(t)\right),      \lb{22}  
\end{equation}
\begin{equation}
    \frac{d\mu^2_{\text{run}}}{dt} = \mu^2_{\text{run}}(t)\beta_{(\mu^2)}
             \left(g_{\text{run}}(t),\,\lambda_{\text{run}}(t)\right),
                                                  \lb{23}      
\end{equation}
\begin{equation}
    \frac{dg^2_{\text{run}}}{dt} = \beta_{g}\left(g_{\text{run}}(t),\,\lambda_{\text{run}}(t)\right).
                                           \lb{24}     
\end{equation}
So far as the mathematical structure of HMM is equivalent
to HSED, we can use all results of the scalar electrodynamics
in our calculations, replacing the electric charge $e$ and photon
field $A_{\mu}$ by magnetic charge $g$ and dual gauge field $B_{\mu}$.

The one-loop results for $\beta_{\lambda}$, $\beta_{(\mu^2)}$,
$\beta_{g}$, and $\gamma$ are given in [18,19] for scalar field
with electric charge $e$. Using these results, we obtain for monopoles with charge $g=g_{\text{run}}$ %
the following expressions in the one-loop approximation:
\begin{equation}
\frac{d\lambda_{\text{run}}}{dt}\approx \beta_{\lambda}^{(1)} = \frac 1{16\pi^2}
( 3g^4_{\text{run}} +10 \lambda^2_{\text{run}} - 6\lambda_{\text{run}}g^2_{\text{run}}),
                                     \lb{31}
\end{equation}
\begin{equation}
\frac{d\mu^2_{\text{run}}}{dt}\approx \beta_{(\mu^2)}^{(1)}
= \frac{\mu^2_{\text{run}}}{16\pi^2}( 4\lambda_{\text{run}} - 3g^2_{\text{run}} ),
                                                \lb{32}
\end{equation}
\begin{equation}
    \frac{dg^2_{\text{run}}}{dt}\approx
     \beta_{g}^{(1)} = \frac{g^4_{\text{run}}}{48\pi^2},  \lb{33}
\end{equation}
\begin{equation}
          \gamma^{(1)} = - \frac{3g_{\text{run}}^2}{16\pi^2}.   \lb{30a}
\end{equation}
The RG $\beta$ functions for different renormalizable gauge theories with
semisimple group have been calculated in the two-loop approximation
[\citen{22}--\citen{27}] and even beyond \ct{28}. But in this paper we made use of
the results of \ct{22} and \ct{25} for calculation of $\beta$ functions
and anomalous dimension in the two-loop approximation, applied to the
HMM with scalar monopole fields.

On the level of two-loop approximation we have for all
$\beta$ functions:
\begin{equation}
  \beta = \beta^{(1)} + \beta^{(2)},           \lb{34}   
\end{equation}
where
\begin{equation}
  \beta_{\lambda}^{(2)} = \frac{1}{(16\pi^2)^2}( - 25\lambda^3 +
   \frac{15}{2}g^2{\lambda}^2 - \frac{229}{12}g^4\lambda - \frac{59}{6}g^6)
                                                    \lb{35}     
\end{equation}
and
\begin{equation}
\beta_{(\mu^2)}^{(2)} = \frac{1}{(16\pi^2)^2}(\frac{31}{12}g^4 + 3\lambda^2).
                                             \lb{36}            
\end{equation}
The gauge coupling $\beta_{g}^{(2)}$ function is given by \ct{22}:
\begin{equation}
     \beta_{g}^{(2)} = \frac{g^6}{(16\pi^2)^2}.  \lb{37}      
\end{equation}
Anomalous dimension follows from the calculations made in \ct{25}:
\begin{equation}
    \gamma^{(2)} = \frac{1}{(16\pi^2)^2}\frac{31}{12}g^4.
                                                   \lb{38}      
\end{equation}
In Eqs.(\ref{34})--(\ref{38}) and below, for simplicity, we have used the
following notations: $\lambda\equiv \lambda_{\text{run}}$, $g\equiv g_{\text{run}}$, and
$\mu\equiv \mu_{\text{run}}$.

\section{The phase diagram in the Higgs monopole model}

Now we  want to apply the effective potential calculation as a
technique for the getting phase diagram information for the condensation
of monopoles in HMM.
As it was mentioned in Section 2,
the effective potential (\ref{20}) can have several minima. Their positions
depend on $g^2$, $\mu^2$, and $\lambda$. If the first local minimum occurs
at $\phi = 0$ and $V_{\text{eff}}(0) = 0$, it corresponds to the Coulomb-like phase.
In the case when the effective potential has the second local minimum at
$\phi = \phi_{\text{min}} \neq 0\,$ with $\,V_{\text{eff}}^{\text{min}}(\phi_{\text{min}}^2) < 0$,
we have the confinement phase. The phase transition between the
Coulomb-like and confinement phases is given by the condition when
the first local minimum at $\phi = 0$ is degenerate with the second minimum
at $\phi = \phi_0$.
These degenerate minima are shown in Fig.2 by the curve {\itshape 1}. They correspond
to the different vacua arising in this model. And the dashed curve {\itshape 2}
describes the appearance of two minima corresponding to the confinement
phases (see details in the next Section).

The conditions of the existence of degenerate vacua are given by the
following equations:
\begin{equation}
           V_{\text{eff}}(0) = V_{\text{eff}}(\phi_0^2) = 0,     \lb{39}   
\end{equation}
\begin{equation}
    \frac{\partial V_{\text{eff}}}{\partial \phi}|_{\phi=0} =
    \frac{\partial V_{\text{eff}}}{\partial \phi}|_{\phi=\phi_0} = 0,
\quad{\mbox{or}}\quad V'_{\text{eff}}(\phi_0^2)\equiv
\frac{\partial V_{\text{eff}}}{\partial \phi^2}|_{\phi=\phi_0} = 0,
                                                    \lb{40}    
\end{equation}
and inequalities
\begin{equation}
    \frac{\partial^2 V_{\text{eff}}}{\partial \phi^2}|_{\phi=0} > 0, \qquad
    \frac{\partial^2 V_{\text{eff}}}{\partial \phi^2}|_{\phi=\phi_0} > 0.
                                               \lb{41}       
\end{equation}
The first equation (\ref{39}) applied to Eq.(\ref{20}) gives:
\begin{equation}
    \mu^2_{\text{run}} = - \frac{1}{2} \lambda_{\text{run}}(t_0)\,\phi_0^2\, G^2(t_0),
\quad{\mbox{where}}\quad t_0 = \log(\phi_0^2/M^2).
                                    \lb{42}     
\end{equation}
Calculating the first derivative of $V_{\text{eff}}$ given by Eq.(\ref{40}),
we obtain the following expression:
\begin{equation} 
      V'_{\text{eff}}(\phi^2) = \frac{V_{\text{eff}}(\phi^2)}{\phi^2}(1 +
 2\frac{d\log G}{dt}) + \frac 12 \frac{d\mu^2_{\text{run}}}{dt} G^2(t)
      + \frac 14 \biggl(\lambda_{\text{run}}(t) + \frac{d\lambda_{\text{run}}}{dt} +
      2\lambda_{\text{run}}\frac{d\log G}{dt}\biggr)G^4(t)\phi^2.
                                                \lb{45}           
\end{equation}
From Eq.(\ref{21}), we have:
\begin{equation}
          \frac{d\log G}{dt} = - \frac{1}{2}\gamma .  \lb{46}        
\end{equation}
It is easy to find the joint solution of equations
\begin{equation}
      V_{\text{eff}}(\phi_0^2) = V'_{\text{eff}}(\phi_0^2) = 0.       \lb{47} 
\end{equation}
Using RGE (\ref{22}), (\ref{23}) and Eqs.(\ref{42})--(\ref{46}), we obtain:
\begin{equation}
 V'_{\text{eff}}(\phi_0^2) =\frac{1}{4}( - \lambda_{\text{run}}\beta_{(\mu^2)} +
\lambda_{\text{run}} + \beta_{\lambda} - \gamma \lambda_{\text{run}})G^4(t_0)\phi_0^2 = 0,
                                                    \lb{48}       
\end{equation}
or
\begin{equation}
    \beta_{\lambda} + \lambda_{\text{run}}(1 - \gamma - \beta_{(\mu^2)}) = 0.
                                            \lb{49}   
\end{equation}
Substituting in Eq.(\ref{49}) the functions
$\beta_{\lambda}^{(1)},\,\beta_{(\mu^2)}^{(1)},$ and $\gamma^{(1)}$
given by Eqs.(\ref{31}), (\ref{32}), and (\ref{30a}), we obtain
in the one-loop approximation the following equation for the phase transition border:
\begin{equation}
     g^4_{\text{PT}} = - 2\lambda_{\text{run}}(\frac{8\pi^2}3 + \lambda_{\text{run}}).
                                                 \lb{50}
\end{equation}
The curve (\ref{50}) is represented on the phase diagram
$(\lambda_{\text{run}}; g^2_{\text{run}})$ of Fig.3 by the curve {\itshape 1} which describes
the border between the "Coulomb-like" phase with $V_{\text{eff}} \ge 0$
and the confinement one with $V_{\text{eff}}^{\text{min}} < 0$. This border corresponds to
the one-loop approximation.

Using Eqs.(\ref{31}), (\ref{32}), (\ref{30a})--(\ref{36}), and (\ref{38}), we are able
to construct the phase transition border in the two-loop approximation.
Substituting these equations into Eq.(\ref{49}), we obtain the following
equation for the phase transition border in the two-loop approximation:
\begin{equation}
 3y^2 - 16\pi^2 + 6x^2 + \frac{1}{16\pi^2}(28x^3 + \frac{15}{2}x^2y +
  \frac{97}{4}xy^2 - \frac{59}{6}y^3) = 0,            \lb{51}   
\end{equation}
where $x = - \lambda_{\text{PT}}$ and $y = g^2_{\text{PT}}$ are the phase transition
values of $ - \lambda_{\text{run}}$ and $g^2_{\text{run}}$.
Choosing the physical branch corresponding to $g^2 \ge 0$ and $g^2\to 0$,
when $\lambda \to 0$, we have received curve {\itshape 2} on the phase diagram
$(\lambda_{\text{run}}; g^2_{\text{run}})$ shown in Fig.3. This curve
corresponds to the two-loop approximation and can be compared with
the  curve {\itshape 1} of Fig.3, which describes the same phase border
calculated in the one-loop approximation.
It is easy to see from the comparison of curves {\itshape 1} and {\itshape 2} that the accuracy of the one-loop
approximation is not excellent and can commit errors of order 30\%.

According to the phase diagram drawn in Fig.3, the confinement phase
begins at $g^2 = g^2_{\text{max}}$ and exists under the phase transition border line
in the region $g^2 \le g^2_{\text{max}}$, where $e^2$ is large:
$e^2\ge (2\pi/g_{\text{max}})^2$ due to the Dirac relation (see below).
Therefore, we have:
$$
g^2_{\text{crit}} = g^2_{\text{max1}}\approx 18.61
\quad -\quad
{\mbox{in the one-loop approximation}},\quad
$$
\begin{equation}
   g^2_{\text{crit}} = g^2_{\text{max2}}\approx
  15.11 \quad -\quad {\mbox{in the two-loop approximation}}.  \lb{52} 
\end{equation}
Comparing these results, we obtain the accuracy of
deviation between them of order 20\%.

The results (\ref{52}) give:
$$
   \tilde \alpha_{\text{crit}} = \frac {g^2_{\text{crit}}}{4\pi}\approx 1.48,
\quad -\quad{\mbox{in the one-loop approximation}},
$$
\begin{equation}
   \tilde \alpha_{\text{crit}} = \frac {g^2_{\text{crit}}}{4\pi}\approx 1.20,
  \quad -\quad {\mbox{in the two-loop approximation}}.
                                                  \lb{53}          
\end{equation}
Using the Dirac relation for elementary charges:
\begin{equation}
   eg = 2\pi, \quad{\mbox{or}}\quad \alpha \tilde \alpha = \frac{1}{4},
                                             \lb{54}         
\end{equation}
we get the following values for the critical electric fine
structure constant:
$$
        \alpha_{\text{crit}} = \frac{1}{4{\tilde \alpha}_{\text{crit}}}\approx 0.17
\quad -\quad{\mbox{in the one-loop approximation}},
$$
\begin{equation}
        \alpha_{\text{crit}} = \frac{1}{4{\tilde \alpha}_{\text{crit}}}\approx 0.208
  \quad -\quad {\mbox{in the two-loop approximation}}.
                                               \lb{55}
\end{equation}
The last result coincides with the lattice values (\ref{4}) obtained for the
compact QED by Monte Carlo method \ct{10s}.

Writing Eq.(\ref{24}) with $\beta_{g}$ function given by Eqs.(\ref{33}),
(\ref{34}), and (\ref{37}), we have the following RGE for the monopole
charge in the two-loop approximation:
\begin{equation}
    \frac{dg^2_{\text{run}}}{dt}\approx
    \frac{g^4_{\text{run}}}{48\pi^2} + \frac{g_{\text{run}}^6}{(16\pi^2)^2},
                                        \lb{56a}
\end{equation}
or
\begin{equation}
    \frac{d\log{\tilde \alpha}}{dt}\approx
          \frac{\tilde \alpha}{12\pi}(1 + 3\frac{\tilde \alpha}{4\pi}).
                                        \lb{56b}
\end{equation}
The values (\ref{52}) for $g^2_{\text{crit}}=g^2_{\text{max}1,2}$ indicate that
the contribution of two loops described by the second term of Eq.~(\ref{56a}) or Eq.~(\ref{56b})
is about 0.3 confirming a validity of the perturbation theory.

In general, we are able to estimate the validity of two-loop approximation
for all $\beta$-functions calculating the corresponding
ratios of two-loop contributions to one-loop contributions
at the maximum of curve {\itshape 2}, where
\begin{equation}
\lambda_{\text{crit}} =
\lambda_{\text{run}}^{\text{max2}}\approx{-7.13}\quad{\mbox{and}}\quad
g^2_{\text{crit}} = g^2_{\text{max2}}\approx{15.11}.                        \lb{max}
\end{equation}
We have the following result:     
\begin{equation}
\begin{array}{l}
\\[-0.2cm]%
\frac{\ds\beta_{(\mu^2)}^{(2)}}{\ds\beta_{(\mu^2)}^{(1)}}\approx{-0.0637},\\[0.8cm]
\frac{\ds\beta_{\lambda}^{(2)}}
{\ds\beta_{\lambda}^{(1)}}\approx{0.0412},\\[0.8cm]
\frac{\ds
\beta_g^{(2)}}{\ds\beta_g^{(1)}}\approx{0.2871}.\\[0.8cm]
\end{array}
                                         \lb{57}
\end{equation}
Here we see that all ratios are sufficiently small, i.e., all
two-loop contributions are small in comparison with one-loop contributions,
confirming the validity of perturbation theory in the two-loop
approximation, considered in this model. The accuracy of deviation is worse
($\sim 30\%$) for $\beta_g$ function. But it is necessary to emphasize
that calculating the border curves {\itshape 1} and {\itshape 2} of Fig.3 we have not used RGE~(\ref{24})
for monopole charge: $\beta_g$ function is absent in Eq.(\ref{49}).
Therefore the calculation of $g^2_{\text{crit}}$ according to Eq.(\ref{51}) does not depend on the approximation of
$\beta_g$ function. The above-mentioned \mbox{$\beta_g$ function} appears only in the second-order
derivative of $V_{\text{eff}}$ which is related with the monopole mass $m$
(see the next section).

Eqs.(\ref{4}) and (\ref{55})
give the following result:
\begin{equation}
          \alpha_{\text{crit}}^{-1}\approx 5.            \lb{56}      
\end{equation}
This value is important for the phase transition at the Planck scale
predicted by MPM.

\section{Triple point}

In this section we demonstrate the existence of the triple point on the
phase diagram of HMM.

Considering the second derivative of the effective potential:
\begin{equation}
                V''_{\text{eff}}(\phi_0^2)
    \equiv \frac{\partial^2 V_{\text{eff}}}{\partial {(\phi^2)}^2},
                                                 \lb{58}     
\end{equation}
we can calculate it for the RG improved effective potential (\ref{20}):
$$
{V''}_{\text{eff}}(\phi^2) = \frac {{V'}_{\text{eff}}(\phi^2)}{\phi^2} + \biggl( - \frac 12
 \mu^2_{\text{run}}
+ \frac 12 \frac{d^2\mu^2_{\text{run}}}{dt^2} + 2\frac{d\mu^2_{\text{run}}}{dt}
 \frac{d\log G}{dt}+
$$
$$
 +  \mu^2_{\text{run}}\frac{d^2\log G}{dt^2} +
   2\mu^2_{\text{run}}{(\frac{d\log G}{dt})}^2\biggl)\frac {G^2}{\phi^2} + \biggl(
   \frac 12 \frac{d\lambda_{\text{run}}}{dt} + \frac 14 \frac {d^2\lambda_{\text{run}}}
  {dt^2} + 2\frac{d\lambda_{\text{run}}}{dt}\frac{d\log G}{dt}+
$$
\begin{equation}
 + 2\lambda_{\text{run}}\frac{d\log G}{dt} + \lambda_{\text{run}}\frac{d^2\log G}{dt^2} +
     4\lambda_{\text{run}}{(\frac{d\log G}{dt})}^2\biggl) G^4(t).
                                                    \lb{59}         
\end{equation}
Let us consider now the case when this second derivative
changes its sign giving a maximum of $V_{\text{eff}}$ instead of the minimum
at $\phi^2 = \phi_0^2$. Such a possibility is shown in Fig.2 by
the dashed curve {\itshape 2}. Now the two additional minima at $\phi^2 = \phi_1^2$
and $\phi^2 = \phi_2^2$ appear in our theory. They correspond to the two
different confinement phases for the confinement of electrically  charged
particles if they exist in the system. When these two minima are degenerate,
we have the following requirements:
\begin{equation}
       V_{\text{eff}}(\phi_1^2) = V_{\text{eff}}(\phi_2^2) < 0\quad     
{\mbox{and}}\quad
        {V'}_{\text{eff}}(\phi_1^2) = {V'}_{\text{eff}}(\phi_2^2) = 0,   \lb{61}  
\end{equation}
which describe the border between the confinement phases conf.1 and conf.2
presented in Fig.4. This border is given as a curve {\itshape 3} at the phase
diagram $(\lambda_{\text{run}}; g^4_{\text{run}})$ drawn in Fig.4. The curve {\itshape 3}
meets the curve {\itshape 1} at the triple point $A$.
According to the illustration shown in Fig.2, it is obvious
that this triple point $A$ is given by the following requirements:
\begin{equation}
    V_{\text{eff}}(\phi_0^2) = V'_{\text{eff}}(\phi_0^2) = V''_{\text{eff}}(\phi_0^2) = 0.
                                         \lb{62}     
\end{equation}
In contrast to the requirements:
\begin{equation}
       V_{\text{eff}}(\phi_0^2) = V'_{\text{eff}}(\phi_0^2) = 0,    \lb{63} 
\end{equation}
giving the curve {\itshape 1}, let us consider now the joint solution of the following
equations:
\begin{equation}
         V_{\text{eff}}(\phi_0^2) = V''_{\text{eff}}(\phi_0^2) = 0 .    \lb{64} 
\end{equation}
For simplicity, we have considered the one-loop approximation.
It is easy to obtain the solution of Eq.(\ref{64}) in the one-loop
approximation, using Eqs.(\ref{59}), (\ref{42}), (\ref{46}), and \mbox{(\ref{31})--(\ref{30a})}:
\begin{equation}
{\cal F}(\lambda_{\text{run}}, g^2_{\text{run}}) = 0,
                                             \lb{65}        
\end{equation}
where
$$
{\cal F}(\lambda_{\text{run}}, g^2_{\text{run}}) = 5g_{\text{run}}^6 +
  24\pi^2g_{\text{run}}^4 + 12\lambda_{\text{run}}g_{\text{run}}^4 - 9\lambda_{\text{run}}^2g_{\text{run}}^2+
$$
\begin{equation}
  + 36\lambda_{\text{run}}^3 + 80\pi^2\lambda_{\text{run}}^2 + 64\pi^4\lambda_{\text{run}}.
                                                \lb{66}     
\end{equation}
The dashed curve {\itshape 2} of Fig.4 represents the solution of
Eq.(\ref{65}) which is equivalent to Eqs.(\ref{64}). The curve {\itshape 2}
is going very close to the maximum of the curve {\itshape 1}. Assuming that the
position of the triple point A coincides with this maximum,
let us consider the border between the phase conf.1, having the first
minimum at nonzero $\phi_1$  with $V_{\text{eff}}^{\text{min}}(\phi_1^2) = c_1 < 0$,
and the phase conf.2 which reveals two minima with the second minimum
being the deeper one and having $V_{\text{eff}}^{\text{min}}(\phi_2^2)=c_2 < 0$.
This border (described by the curve {\itshape 3} of Fig.4) was calculated in the
vicinity of the triple point $A$ by means of Eqs.(\ref{61})
with $\phi_1$ and $\phi_2$ represented as $ \phi_{1,2} = \phi_0 \pm \epsilon$
with $\epsilon\ll\phi_0$. The result of such calculations gives the
following expression for the curve {\itshape 3}:
\begin{equation}
  g^4_{\text{PT,3}} = \frac {5}{2} ( 5\lambda_{\text{run}} + 8 \pi^2) \lambda_{\text{run}} + 8\pi^4.
                                                 \lb{67}         
\end{equation}

The curve {\itshape 3} meets the curve {\itshape 1} at the triple point $A$.

The piece of the curve {\itshape 1} to the left of the point $A$ describes the border between the Coulomb-like phase and
phase conf.1. In the vicinity of the triple point $A$ the second derivative $V_{\text{eff}}''(\phi_0^2)$ changes
its sign leading to the existence of the maximum at $\phi^2=\phi_0^2$, in correspondence with the dashed curve {\itshape 2}
of Fig.2.  By this reason, the curve {\itshape 1} of Fig.4 does not describe a phase transition border from the point $A$
to the point B when the curve {\itshape 2} again intersects the curve {\itshape 1} at $\lambda_{\text{(B)}}\approx - 12.24$. This
intersection (again giving $V''_{\text{eff}}(\phi_0^2) > 0$) occurs surprisingly quickly.

The right piece of the curve {\itshape 1} to the right of the point $B$
separates the Coulomb-like phase and the phase "conf.2". But
between the points A and B the phase transition border is going
slightly above the curve {\itshape 1}. This deviation is very small and cannot be
distinguished on Fig.4.

It is necessary to note that only $V''_{\text{eff}}(\phi^2)$ contains
the derivative $dg^2_{\text{run}}/dt$.
The joint solution of equations (\ref{62})
leads to the joint solution of Eqs.(\ref{50}) and (\ref{65}).
This solution was obtained numerically and gave the following triple point
values of $\lambda_{\text{run}}$ and $g^2_{\text{run}}$:
\begin{equation}
    \lambda_{\text{(A)}}\approx{ - 13.4073},\quad
              g^2_{\text{(A)}}\approx{18.6070}.            \lb{68}  
\end{equation}
The solution ({\ref{68}}) demonstrates that the triple point $A$
exists in the very neighborhood of maximum of the curve (\ref{50}).
The position of this maximum is given by the following analytical
expressions, together with their approximate values:
\begin{equation}
     \lambda_{\text{(A)}}\approx - \frac{4\pi^2}3\approx -13.2,
                                                       \lb{69}     
\end{equation}
\begin{equation}
     g^2_{\text{(A)}} = g^2_{\text{crit}}|_{\mbox{for}\;\lambda_{\text{run}}=\lambda_{\text{(A)}}}
                 \approx \frac{4\sqrt{2}}3{\pi^2}\approx 18.6.
                                                        \lb{70}     
\end{equation}
Finally, we can conclude that the phase diagram shown in Fig.4 gives
such a description: there exist three phases in the dual sector of the Higgs
scalar electrodynamics -- the Coulomb-like phase and confinement
phases conf.1 and conf.2.

The border {\itshape 1}, which is described by the curve (\ref{50}), separates
the Coulomb-like phase (with $V_{\text{eff}} \ge 0$) and confinement phases
(with $V_{\text{eff}}^{\text{min}}(\phi_0^2) < 0$).
The curve {\itshape 1} corresponds to the joint solution of the equations
$V_{\text{eff}}(\phi_0^2)=V'_{\text{eff}}(\phi_0^2)=0$.

The dashed curve {\itshape 2} represents the solution of the equations
$V_{\text{eff}}(\phi_0^2)=V''_{\text{eff}}(\phi_0^2)=0$.

The phase border {\itshape 3} of Fig.4 separates two confinement phases.
The following requirements take place for this border:
$$
          V_{\text{eff}}(\phi_{1,2}^2) < 0,\qquad
         V_{\text{eff}}(\phi_1^2) = V_{\text{eff}}(\phi_2^2),\qquad
         V'_{\text{eff}}(\phi_1^2) = V'_{\text{eff}}(\phi_2^2) = 0,
$$
\begin{equation}
         V''_{\text{eff}}(\phi_1^2) > 0,\qquad V''_{\text{eff}}(\phi_2^2) > 0.
                                               \lb{71}    
\end{equation}
The triple point $A$ is a boundary point of all three phase transitions
shown in the phase diagram of Fig.4.
For $g^2 < g^2_{\text{({A})}}$ the field system described by our model exists
in the confinement phase, where all electric charges  have to be confined.

Taking into account that monopole mass $m$ is given by the
following expression:
\begin{equation}
  V''_{\text{eff}}(\phi_0^2) =
\frac {1}{4\phi_0^2}\frac {d^2V_{\text{eff}}}{d\phi^2}|_{\phi=\phi_0}
            = \frac {m^2}{4\phi_0^2},                \lb{72}
\end{equation}
we see that monopoles acquire zero mass in the vicinity of the triple point $A$:
\begin{equation}
  V''_{\text{eff}}(\phi_{\text{0A}}^2) = \frac {m^2_{\text{(A)}}}{4\phi_{\text{0A}}^2} = 0.
                                                     \lb{74}
\end{equation}
This result is in agreement with the result of compact QED \ct{29}:
$m^2\to 0$ in the vicinity of the critical point.

\section {"ANO-strings", or the vortex description of the
confinement phases}

As it was shown in the previous Section, two regions between the curves
{\itshape 1}, {\itshape 3} and {\itshape 3}, {\itshape 1}, given by the phase diagram of Fig.4, correspond to the
existence of the two confinement phases, different in the sense
that the phase conf.1 is produced by the second minimum, but the phase
conf.2 corresponds to the third minimum of the effective potential.
It is obvious that in our case both phases have nonzero monopole condensate
in the minima of the effective potential, when
$V_{\text{eff}}^{\text{min}}(\phi_{1,2}\neq 0) < 0$. By this reason, the
Abrikosov--Nielsen--Olesen (ANO) electric vortices
(see \ct{30,31}) may be created in both these phases.
Only closed strings exist in the confinement phases of HMM.
The properties of ANO-strings in the $U(1)$ gauge theory were
investigated in \ct{32}.

\section{Multiple Point Model and critical values
of the {\bfseries{\itshape{U(1)}}} and {\bfseries{\itshape{SU(N)}}} fine structure constants}

\subsection{Anti-grand unification theory}

Grand Unification Theories (GUTs) were constructed with aim to extend
the Standard Model (SM).
The supersymmetric extension of the SM consists of taking the
SM and adding the corresponding supersymmetric partners \ct{32a}.  The Minimal
Supersymmetric Standard Model (MSSM) shows the possibility of the existence of
the grand unification point at $\mu_{\text{GUT}}\sim 10^{16}$ GeV \ct{33a}.
Unfortunately, at present time
experiment does not indicate any manifestation of the supersymmetry.
In this connection, the Anti-Grand Unification Theory (AGUT) was developed
in [\citen{33}--\citen{38}] as a realistic alternative to SUSY GUTs.
According to this theory, supersymmetry does not come into the existence
up to the Planck energy scale (\ref{1a}).

The SM is based on the group:
\begin{equation}
   SMG = SU(3)_c \times SU(2)_L \times U(1)_Y.    \lb{75}  
\end{equation}
AGUT suggests that at the scale $\mu_{\text{G}}\sim \mu_{\text{Pl}}=M_{\text{Pl}}$
there exists the more fundamental group $G$ containing $N_{\text{gen}}$
copies of the Standard Model Group (SMG):
\begin{equation}
G = SMG_1\times SMG_2\times...\times SMG_{N_{\text{gen}}}\equiv (SMG)^{N_{\text{gen}}},
                                                  \lb{76}   
\end{equation}
where $N_{\text{gen}}$ designates the number of quark and lepton generations.

If $N_{\text{gen}}=3$ (as AGUT predicts), then the fundamental gauge group $G$ is:
\begin{equation}
    G = (SMG)^3 = SMG_{\text{1st gen}}\times SMG_{\text{2nd gen}}\times SMG_{\text{3rd gen}},
                                        \lb{77} 
\end{equation}
or the generalized one:
\begin{equation}
         G_{\text{f}} = (SMG)^3\times U(1)_{\text{f}},           \lb{78}  
\end{equation}
which was suggested by the fitting of fermion masses of the SM
(see \ct{35}).

Recently a new generalization of AGUT was suggested in \ct{37}:
\begin{equation}
           G_{\text{ext}} = (SMG\times U(1)_{\text{B-L}})^3,    \lb{79}  
\end{equation}
which takes into account the see-saw mechanism with right-handed neutrinos,
also gives the reasonable fitting of the SM fermion masses and describes
all neutrino experiments known today.

The group $G_{\text{f}}$ contains the following gauge fields:
$3\times 8 = 24$ gluons, $3\times 3 = 9$ $W$ bosons, and $3\times 1 + 1 = 4$
Abelian gauge bosons.

At first sight, this ${(SMG)}^3\times U(1)_{\text{f}}$ group with its 37 generators
seems to be just one among many possible SM gauge group extensions.
However, it is not such an arbitrary choice. There are reasonable requirements (postulates) on the gauge group
$G$ (or $G_{\text{f}}$, or $G_{\text{ext}}$) which unambiguously specify this group.  It should obey the
following postulates (the first two are also valid for $SU(5)$ GUT):

\vspace{0.1cm}

1. $G$ or $G_{\text{f}}$ should only contain transformations, transforming the
known 45 Weyl fermions ( = 3 generations of 15 Weyl particles each)
-- counted as left-handed, say -- into each other unitarily,
so that $G$ (or $G_{\text{f}}$) must be a subgroup of $U(45)$: $G\subseteq U(45)$.

\vspace{0.1cm}

2. No anomalies, neither gauge nor mixed. AGUT assumes that only
straightforward anomaly cancellation takes place and forbids the
Green--Schwarz type anomaly cancellation \ct{39}.

\vspace{0.1cm}

3. AGUT should NOT UNIFY the irreducible representations under the SM
gauge group, called here SMG (see Eq.(\ref{75})).

\vspace{0.1cm}

4. $G$ is the maximal group satisfying the above-mentioned postulates.

\vspace{0.1cm}

There are five Higgs fields in the extended AGUT with the group of symmetry
$G_{\text{f}}$ \ct{35}. These fields break AGUT to the SM what
means that their vacuum expectation values (VEVs) are active.
The extended AGUT with the group of symmetry $G_{\text{ext}}$ given by
Eq.(\ref{79}) was suggested in \ct{37} with aim to explain the neutrino
oscillations. Introducing the right-handed neutrino in the model, the
authors of this theory replaced the postulate {\itshape 1} and considered $U(48)$ group
instead of $U(45)$, so that $G_{\text{ext}}$ is a subgroup of $U(48)$:  $G_{\text{ext}}\subseteq
U(48)$. This group ends up having $7$ Higgs fields (see details in \ct{37}).
Typical fit to the masses and mixing angles for the SM leptons and quarks in
the framework of the $G_{\text{ext}}$ theory has shown that, in contrast to the old
extended AGUT with the group of symmetry $G_{\text{f}}$, new results are more
encouraging.

\subsection{AGUT-MPM prediction of the Planck scale values of the
{\bfseries{\itshape{U(1)}}}, {\bfseries{\itshape{SU(2)}}}, and {\bfseries{\itshape{SU(3)}}} fine structure constants}

As it was mentioned in Introduction, the AGUT approach is used in
conjunction with MPM [1--4], which assumes the existence
of the Multiple Critical Point (MCP) at the Planck scale.

The usual definition of the SM coupling constants:
\begin{equation}
  \alpha_1 = \frac{5}{3}\frac{\alpha}{\cos^2\theta_{\ov{\text{MS}}}},\quad
  \alpha_2 = \frac{\alpha}{\sin^2\theta_{\ov{\text{MS}}}},\quad
  \alpha_3 \equiv \alpha_{\text{s}} = \frac {g^2_{\text{s}}}{4\pi},     \lb{81}  
\end{equation}
where $\alpha$ and $\alpha_s$ are the electromagnetic and SU(3)
fine structure constants, respectively, is given in the Modified
minimal subtraction scheme ($\ov{MS}$). Here $\theta_{\overline{\text{MS}}}$ is the Weinberg weak angle in $\ov{MS}$ scheme.
Using RGE with experimentally
established parameters, it is possible to extrapolate the experimental
values of three inverse running constants $\alpha_i^{-1}(\mu)$
(here $\mu$ is an energy scale and $i=1,2,3$ correspond to $U(1)$,
$SU(2)$ and $SU(3)$ groups of the SM) from the Electroweak scale to the Planck
scale. The precision of the LEP data allows to make this extrapolation
with small errors (see \ct{33a}). Assuming that these RGEs for
$\alpha_i^{-1}(\mu)$ contain only the contributions of the SM particles
up to $\mu\approx \mu_{\text{Pl}}$ and doing the extrapolation with one
Higgs doublet under the assumption of a "desert", the following results
for the inverses $\alpha_{\text{Y,2,3}}^{-1}$ (here $\alpha_{\text{Y}}\equiv \frac{3}{5}
\alpha_1$) were obtained in \ct{1y} (compare with \ct{33a}):
\begin{equation}
\alpha_{\text{Y}}^{-1}(\mu_{\text{Pl}})\approx 55.5; \quad
\alpha_2^{-1}(\mu_{\text{Pl}})\approx 49.5; \quad
\alpha_3^{-1}(\mu_{\text{Pl}})\approx 54.0.
\lb{82}   
\end{equation}
The extrapolation of $\alpha_{\text{Y,2,3}}^{-1}(\mu)$ up to the point
$\mu=\mu_{\text{Pl}}$ is shown in Fig.5.

According to the AGUT, at some point $\mu=\mu_G < \mu_{\text{Pl}}$ (but near
$\mu_{\text{Pl}}$) the fundamental group $G$ (or $G_f$, or $G_{\text{ext}}$)
undergoes spontaneous breakdown to the diagonal subgroup:
\begin{equation}
      G \longrightarrow G_{\text{diag.subgr.}} = \{g,g,g || g\in SMG\},
                                                          \lb{83}   
\end{equation}
which is identified with the usual (low-energy) group SMG.
The point $\mu_{\text{G}}\sim 10^{18}$ GeV also is shown in Fig.5, together with
a region of $G$ theory where the AGUT works.

The AGUT prediction of the values of $\alpha_i(\mu)$ at $\mu=\mu_{\text{Pl}}$
is based on the MPM assumption about the existence of the phase
transition boundary point MCP at the Planck scale, and gives these values
in terms of the corresponding critical couplings $\alpha_{i,\text{crit}}$
\ct{1y,33,34}:
\begin{equation}
            \alpha_i(\mu_{\text{Pl}}) = \frac {\alpha_{i,\text{crit}}}{N_{\text{gen}}}
                       = \frac{\alpha_{i,\text{crit}}}{3}
                \quad{\mbox{for}}\quad i=2,3,       \lb{84}   
\end{equation}
and
\begin{equation}
\alpha_1(\mu_{\text{Pl}}) = \frac{2\alpha_{1,\text{crit}}}{N_{\text{gen}}(N_{\text{gen}} + 1)}
= \frac{\alpha_{1,\text{crit}}}{6} \quad{\mbox{for}}\quad U(1).
\lb{85}    
\end{equation}
There exists a simple explanation of the relations (\ref{84}) and (\ref{85}).
As it was mentioned above, the group G breaks down at $\mu=\mu_{\text{G}}$.
It should be said that at
the very high energies $\mu_{\text{G}} \le \mu \le\mu_{\text{Pl}}$ (see Fig.5)
each generation has its own gluons, own $W$'s, etc. The breaking makes
only linear combination of a certain color combination of gluons which
exists below $\mu=\mu_{\text{G}}$ and down to the low energies.
We can say that the phenomenological gluon is a linear
combination (with amplitude $1/{\sqrt 3}$ for $N_{\text{gen}}=3$) for each of the
AGUT gluons of the same color combination. This means that coupling constant
for the phenomenological gluon has a strength that is ${\sqrt 3}$ times smaller,
if as we effectively assume that three AGUT $SU(3)$ couplings are equal
to each other.
Then we have the following formula connecting the fine structure constants
of $G$ theory (e.g., AGUT) and low-energy surviving diagonal subgroup
$G_{\text{diag.subg.}}\subseteq {(SMG)}^3$ given by Eq.(\ref{83}):
\begin{equation}
\alpha_{\text{diag},i}^{-1} = \alpha_{\text{1st gen},i}^ {-1} +
\alpha_{\text{2nd gen},i}^{-1} + \alpha_{\text{3rd gen},i}^{-1}.
\lb{86}     
\end{equation}
Here $i = U(1),\;SU(2),\;SU(3)$, and $i=3$ means that we talk about the gluon
couplings.
For non-Abelian theories we immediately obtain Eq.(\ref{84}) from
Eq.(\ref{86}) at the critical point (MCP).

In contrast to non-Abelian theories, in which the gauge invariance
forbids the mixed (in generations) terms in the Lagrangian of
$G$ theory, the $U(1)$ sector of AGUT contains such mixed
terms:
\begin{equation}
\frac{1}{g^2}\sum_{p,q} F_{\mu\nu,\; p}F_{q}^{\mu\nu} =
\frac{1}{g^2_{11}}F_{\mu\nu,\; 1}F_{1}^{\mu\nu} +
\frac{1}{g^2_{12}}F_{\mu\nu,\; 1}F_{2}^{\mu\nu} +
...
+ \frac{1}{g^2_{23}}F_{\mu\nu,\; 2}F_{3}^{\mu\nu} +
\frac{1}{g^2_{33}}F_{\mu\nu,\; 3}F_{3}^{\mu\nu},
\lb{87}
\end{equation}
where $p,q = 1,2,3$ are the indices of three generations of the AGUT
group $(SMG)^3$. Eq.(\ref{87}) explains the difference between the expressions
(\ref{84}) and (\ref{85}).

It was assumed in \ct{1y} that the MCP values
$\alpha_{i,\text{crit}}$ in Eqs.(\ref{84}) and (\ref{85}) coincide with (or are very
close to) the triple point values of the effective fine structure
constants given by the generalized lattice $SU(3)$, $SU(2)$, and $U(1)$ gauge
theories [11--14] described by Eqs.(\ref{2}) and (\ref{3}).
Also the authors of~[1] have used an assumption that the effective $\alpha_{\text{crit}}$
does not change its value (at least too much) along the whole borderline
{\itshape 3} of Fig.1 for the phase transition Coulomb-confinement (see details in~[1]).

\subsection{Multiple Point Model and the behavior of the electric fine
structure constant near the phase transition point}

The authors of [\citen{7s}--\citen{10s}] were not able to obtain the lattice
triple point values of $\alpha_{i,\text{crit}}$ by Monte Carlo simulations
method. Only the critical value of the electric fine structure
constant $\alpha$ was obtained in \ct{10s} in the compact QED described
by the simple Wilson action corresponding to the case $\gamma^{\text{lat}}=0$
in Eq.(\ref{3}). The result of \ct{10s} for the behavior of $\alpha(\beta)$
in the vicinity of the phase transition point $\beta_{\text{T}}$ is shown in Fig.6(a)
for the Wilson and Villain lattice actions.
Here $\beta\equiv \beta^{\text{lat}}=1/e_0^2$ and $e_0$ is the bare electric charge.
The Villain lattice action is:
\begin{equation}
S_{\text V} =  (\beta/2))\sum_{\text{p}} {(\Theta_{\text{p}} - 2\pi k)}^2, \qquad k\in Z.
\lb{87a}
\end{equation}
Fig.6(b) demonstrates the comparison of the functions
$\alpha(\beta)$ obtained by Monte Carlo method for the Wilson
lattice action and by theoretical calculation of the same quantity.
The theoretical (dashed) curve was calculated by so-called Parisi improvement
formula \ct{40}:
\begin{equation}
\alpha (\beta )=[4\pi \beta W_{\text{p}}]^{-1}.
\lb{88}
\end{equation}
Here $W_{\text{p}}=<\cos \Theta_{\text{p}} >$ is a mean value of the plaquette energy.
The corresponding values of $W_{\text{p}}$ are taken from \ct{9s}.

According to Fig.6(b):
\begin{equation}
            \alpha_{\text{crit,theor}}^{-1}\approx 8.         \lb{89}
\end{equation}
This result does not coincide with the lattice and HMM result (\ref{56}).
The deviation of theoretical calculations from the lattice ones has
the following explanation: Parisi improvement formula (\ref{88})
is valid in Coulomb phase where the mass of artifact monopoles is infinitely
large and the photon is massless. But in the vicinity of the phase
transition (critical) point the monopole mass $m\to 0$ and the photon
acquires the non-zero mass $m_0\neq 0$. This phenomenon leads to
the "freezing" of $\alpha$ at the phase transition point:
the effective electric fine structure constant is almost unchanged
in the confinement phase and approaches its maximal value
$\alpha=\alpha_{\text{max}}$. The authors of \ct{41} predicted $\alpha_{\text{max}}
= \frac{\ds\pi}{\ds12}\approx 0.26$ due to the Casimir effect (see also \ct{3y}).
The analogous freezing of $\alpha_{\text{s}}$ was considered in \ct{42}
in QCD. We also see that Fig.6(a) demonstrates the tendency to the freezing
of $\alpha$.

Now let us consider $\alpha_{\text{Y}}^{-1}\,(\approx \alpha^{-1})$ at the point
$\mu=\mu_{\text{G}}$ shown in Fig.5. If the point $\mu=\mu_{\text{G}}$ is very close
to the Planck scale $\mu=\mu_{\text{Pl}}$, then according to Eqs.(\ref{82}) and
(\ref{85}), we have:
\begin{equation}
\alpha_{\text{1st gen}}^{-1}\approx \alpha_{\text{2nd gen}}^{-1}\approx
\alpha_{\text{3rd gen}}^{-1}\approx \frac{\alpha_{\text{Y}}^{-1}(\mu_{\text{G}})}{6}\approx 9,
\lb{90}
\end{equation}
what is very close to the value (\ref{89}). This means (see Fig.6(b))
that in the $U(1)$ sector of $G$ theory we have $\alpha$ near the critical
point, therefore we can expect the existence of MCP at the Planck scale.
As a consequence of such a prediction, we have to expect the change of
the evolution of $\alpha_i^{-1}(\mu)$ in the region $\mu > \mu_{\text{G}}$ shown in
Fig.5 by dashed lines. Instead of these dashed lines, we have to see the
decreasing of $\alpha_i^{-1}(\mu)$ approaching to MCP at the Planck scale,
where $\alpha_{\text{crit}}$ is close to the value (\ref{56}) obtained in the
present paper. But this is an aim of our future investigations based
on the idea that MCP rules over the evolution of all fine structure
constants in the SM and beyond it.

\section{Conclusions}

In the present paper we have considered the dual Abelian Higgs model
of scalar monopoles reproducing a confinement mechanism in the
lattice gauge theories. Using the Coleman--Weinberg idea of the RG
improvement of the effective potential \ct{20s}, we have considered this
potential with $\beta$-functions calculated in the two-loop approximation.
The phase transition between the Coulomb-like and confinement phases
has been investigated in the $U(1)$ gauge theory by the method developed
in MPM where degenerate vacua are considered. The comparison of the
result $\alpha_{\text{crit}}\approx 0.17$ and ${\tilde \alpha}_{\text{crit}}\approx 1.48$
obtained in the one-loop approximation with the result
$\alpha_{\text{crit}}\approx 0.208$ and ${\tilde \alpha}_{\text{crit}}\approx 1.20$
obtained in the two-loop approximation demonstrates the coincidence of
the critical values of electric and magnetic fine structure constants
calculated in the two-loop approximation of HMM
with the lattice result~\cite{10s}: $\alpha_{\text{crit}}^{\text{lat}}\approx 0.20\pm 0.015$
and ${\tilde \alpha}_{\text{crit}}^{\text{lat}}\approx 1.25\pm 0.10$.
Also comparing the one-loop and two-loop contributions to $\beta$ functions,
we have demonstrated the validity of perturbation theory in solution of the
phase transition problem in the $U(1)$ gauge theory.

In the second part of our paper we have compared the prediction
of AGUT and MPM for the Planck scale values of $\alpha_i^{-1}(\mu)$
with the lattice and HMM results. Such a comparison is very encouraging
for MPM.

\newpage
\clearpage

{\Large\bf Figure captions}

\vspace{1cm}\noindent{\bf Figure 1.} The phase diagram for $U(1)$ when the two-parameter
lattice action is used. This type of action makes it possible to provoke
the confinement $Z_2$ (or $Z_3$) alone. The diagram shows the
existence of a triple (critical) point.
From this triple point emanate three phase borders:
the phase border {\itshape 1} separates the totally confining phase from the phase
where only the discrete subgroup $Z_{2}$ is confined; the phase border
{\itshape 2} separates the latter phase from the totally Coulomb-like phase;
and the phase border {\itshape 3} separates the totally confining and totally
Coulomb-like phases.\\

\noindent{\bf Figure 2.} The effective potential $V_{\text{eff}}$: the curve {\itshape 1} corresponds to
the Coulomb--confinement phase transition; curve {\itshape 2} describes the existence
of two minima corresponding to the confinement phases.\\

\noindent {\bf Figure 3.} The one-loop (curve {\itshape 1}) and two-loop (curve {\itshape 2}) approximation
phase diagram in the dual Abelian Higgs model of scalar monopoles.\\

\noindent {\bf Figure 4.} The phase diagram ($\lambda_{\text{run}};\; {\rm g}^4\equiv {\rm g}^4_{\text{run}}$),
corresponding to the Higgs monopole model in the one-loop approximation,
shows the existence of a triple point A
$\bigl(\lambda_{\text{(A)}} \approx -13.2;\;{\rm g}^2_{\text{(A)}}\approx 18.6\bigr)$.
This triple point is a boundary point of three phase
transitions: the Coulomb-like phase and two confinement phases
(conf.1 and conf.2) meet together at the triple point $A$.
The dashed curve {\itshape 2} shows the requirement:
$V_{\text{eff}}(\phi_0^2) = V''_{\text{eff}}(\phi_0^2) = 0$. Monopole condensation
leads to the confinement of the electric charges: ANO electric
vortices are created in the confinement phases conf.1 and conf.2.\\

\noindent {\bf Figure 5.} The evolution of three inverse running constants
$\alpha_i^{-1}(\mu)$, where $i=1,2,3$ correspond to $U(1)$, $SU(2)$, and $SU(3)$
groups of the SM. The extrapolation of their experimental values
from the electroweak scale to the Planck scale was obtained by using
the renormalization group equations with one Higgs doublet under the
assumption of a "desert". The precision of the LEP data allows to make
this extrapolation with small errors (see \ct{33a}). AGUT works in the region $\mu_{\text{G}}\leq\mu\leq\mu_{\text{Pl}}$.\\

\noindent {\bf Figure 6.} (a) The renormalized electric fine structure constant plotted
versus $\beta /{\beta_{\text{T}}}$ for the Villain action (circles) and the Wilson
action (crosses). The points are obtained in \ct{10s} by the Monte Carlo
simulations method for the compact QED;\\
(b)  The behavior of the effective electric fine structure constant
in the vicinity of the phase transition point obtained with the lattice
Wilson action. The dashed curve corresponds to the theoretical calculations
by the Parisi improvement formula \ct{40}.

\newpage
\clearpage

\noindent
\includegraphics[width=159mm, keepaspectratio=true]{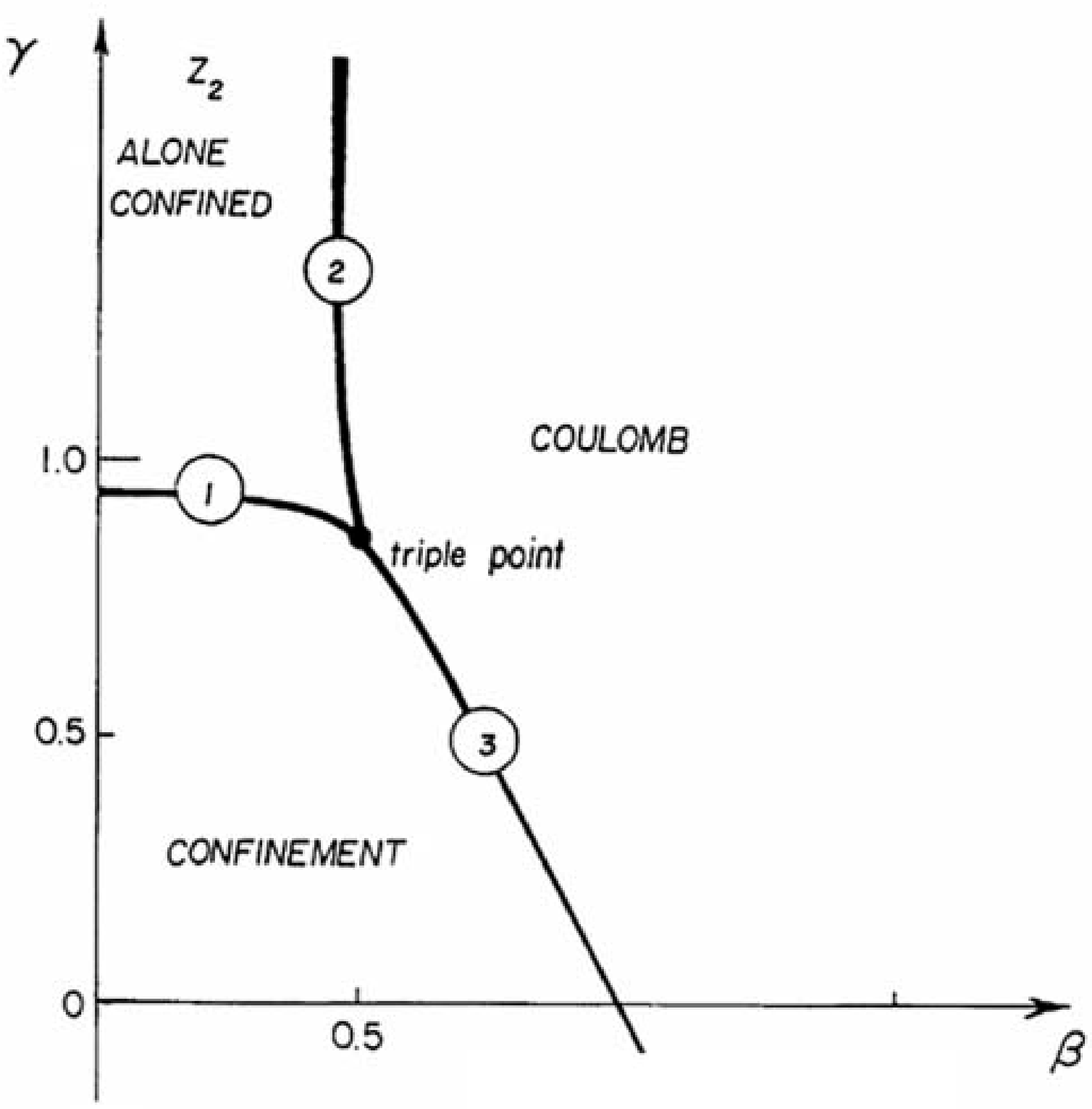}

\vspace*{-5.5cm}\bc{\noindent\Large\bf Fig.1.}\ec

\newpage
\clearpage

\noindent
\includegraphics[width=159mm, keepaspectratio=true]{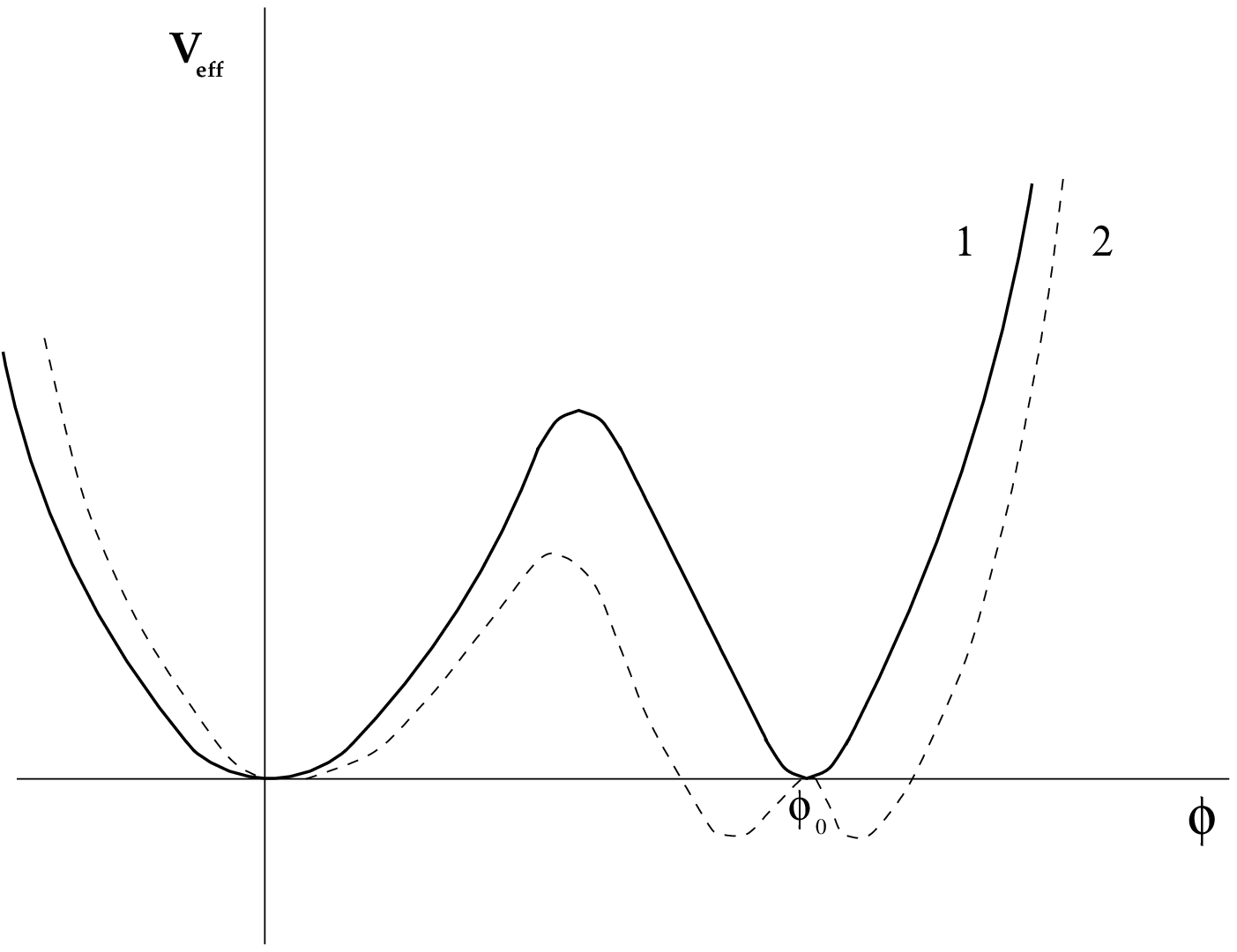}

\vspace*{-2cm}
\bc{\noindent\Large\bf Fig.2.}\ec

\newpage
\clearpage

\noindent\hspace*{-0.75pt}%
\includegraphics[width=159mm, keepaspectratio=true]{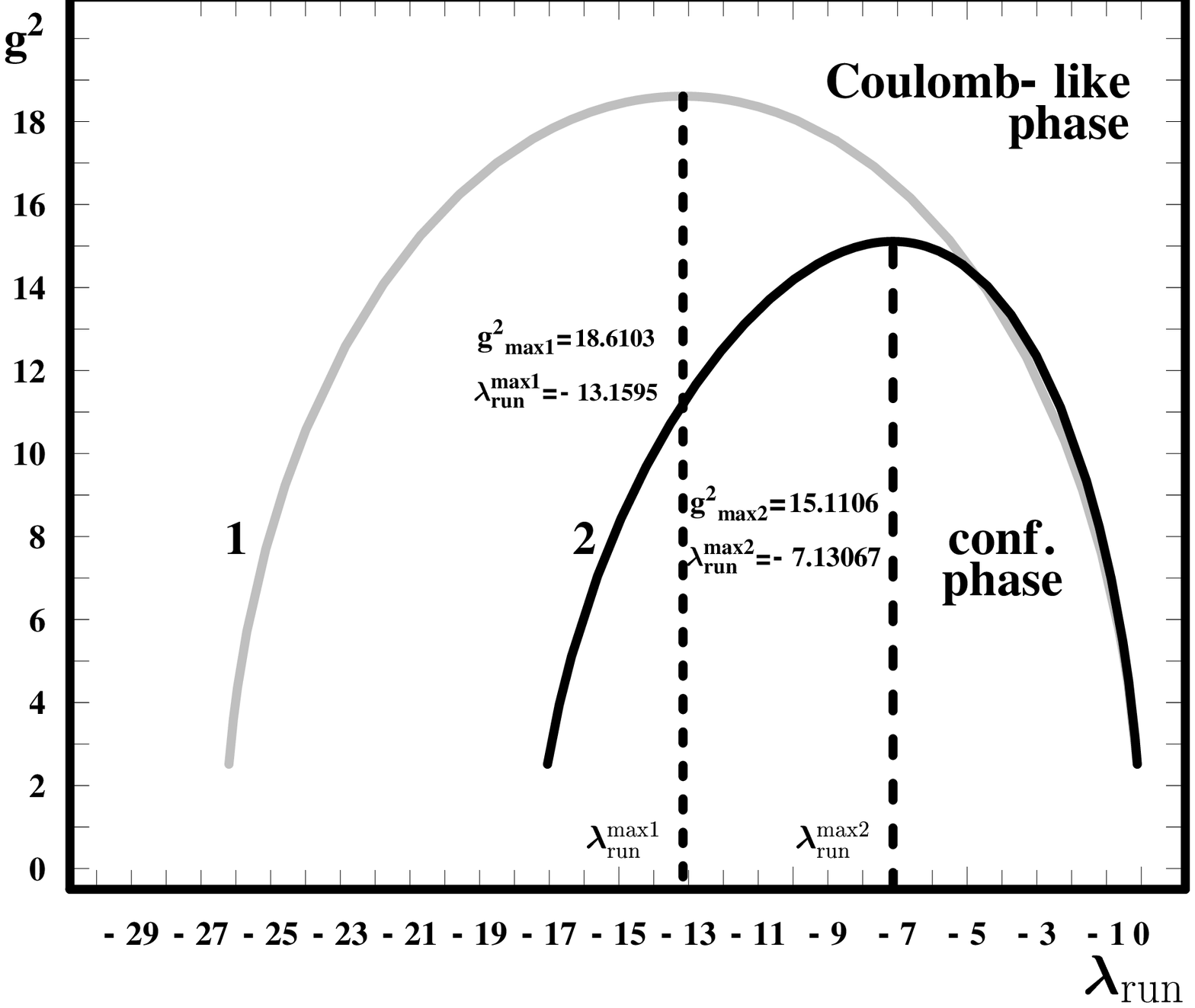}

\vspace*{0.5cm}\bc{\noindent\Large\bf Fig.3.}\ec

\newpage
\clearpage

\noindent
\includegraphics[width=159mm, keepaspectratio=true]{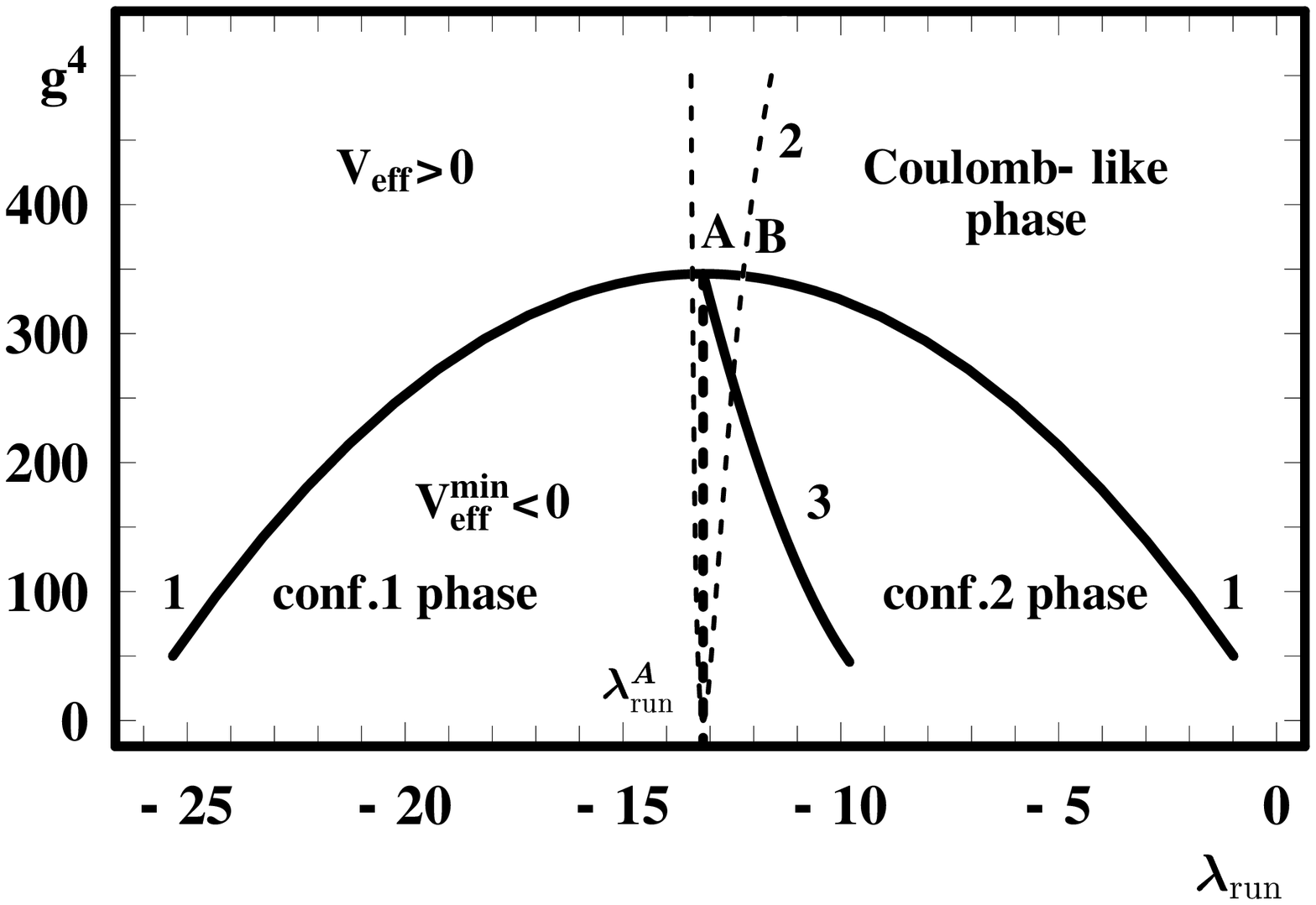}

\vspace*{0.5cm}\bc{\noindent\Large\bf Fig.4.}\ec

\newpage
\clearpage

\noindent
\includegraphics[width=159mm, keepaspectratio=true]{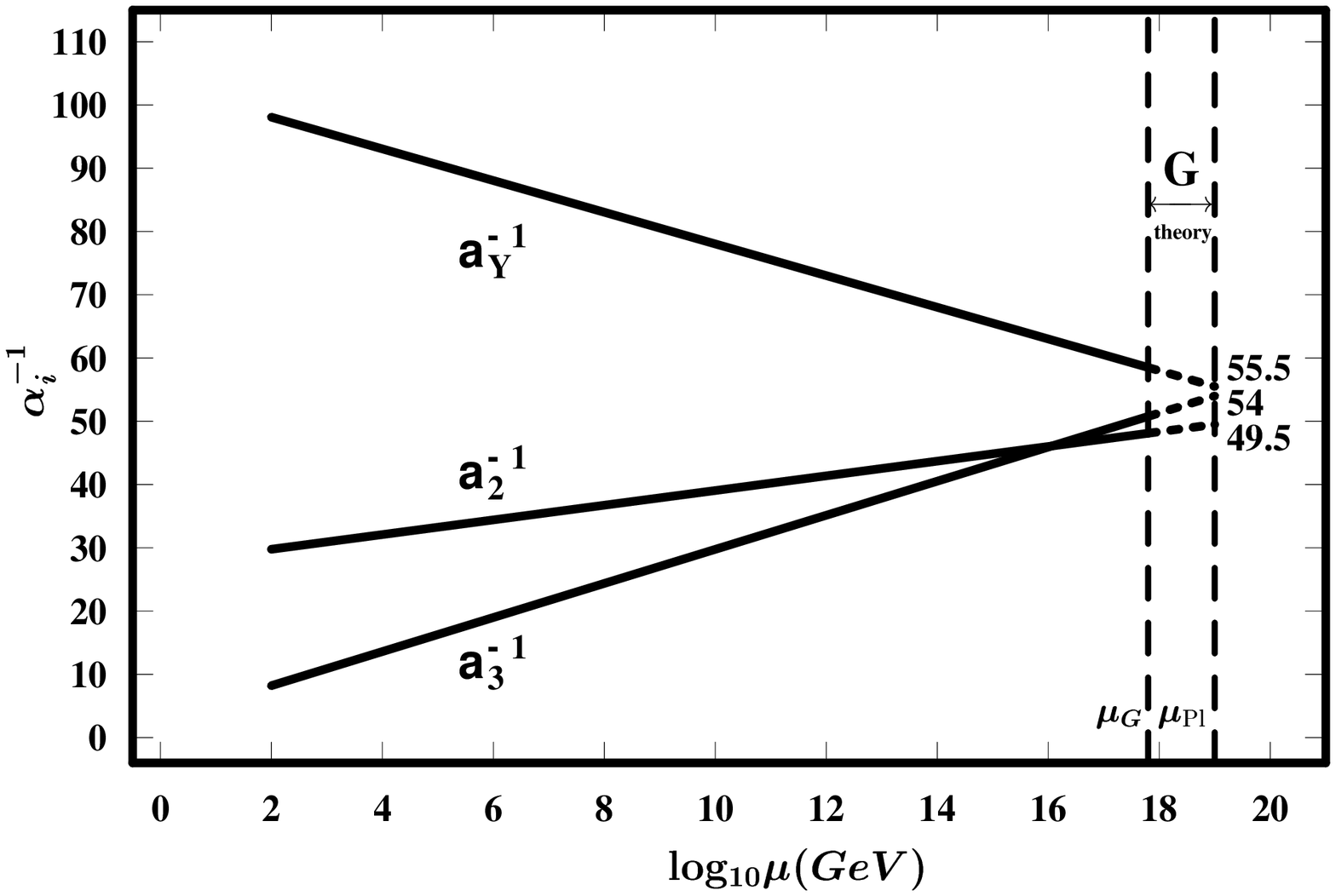}

\vspace*{0.5cm}\bc{\noindent\Large\bf Fig.5.}\ec

\newpage
\clearpage

\noindent\hspace*{-3mm}\includegraphics[width=159mm, keepaspectratio=true]{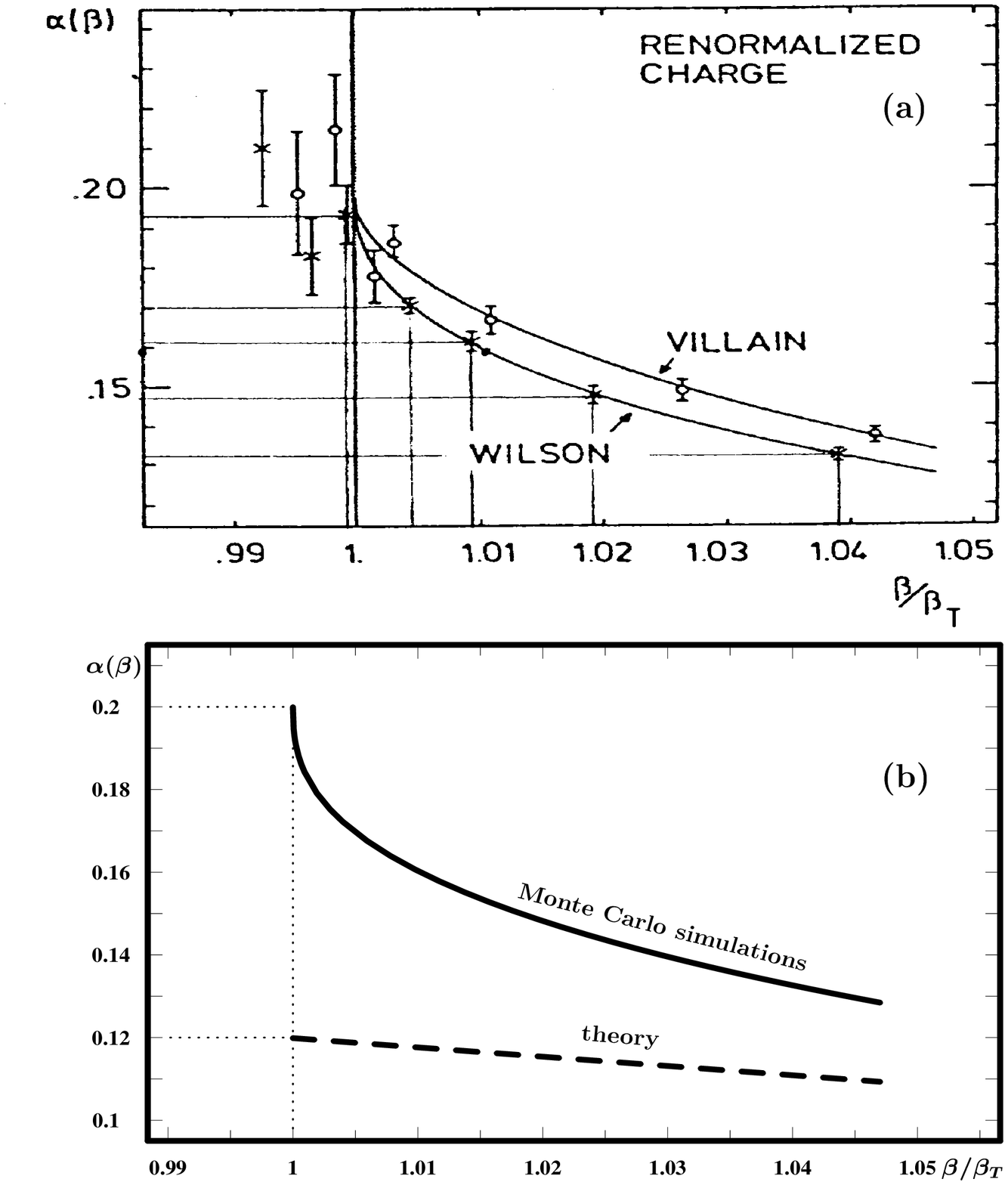}

\vspace*{0.5cm}\bc{\noindent\Large\bf Fig.6(a,b).}\ec

\end{document}